\documentclass[dvips]{imsart}

\RequirePackage[OT1]{fontenc}
\RequirePackage{amsthm,amsmath}
\RequirePackage[numbers]{natbib}
\RequirePackage[colorlinks,citecolor=blue,urlcolor=blue]{hyperref}
\usepackage{graphicx}
\usepackage{amssymb}
\usepackage{epstopdf}

\startlocaldefs

\numberwithin{equation}{section}
\theoremstyle{plain}
\newtheorem{example}{Example}

\begin{document}

%


\title{Logistic regression geometry}

\runtitle{Logistic regression geometry}

\author{\fnms{Karim} \snm{Anaya-Izquierdo}\ead[label=e1]{karim.anaya@lshtm.ac.uk}}
\address{London School of Hygiene and Tropical Medicine,   London   WC1E 7HT, UK \printead{e1}}

\and
\author{\fnms{Frank} \snm{Critchley}\ead[label=e2]{f.critchley@open.ac.uk}}
\address{Department of Mathematics and Statistics,          The Open University,  Milton Keynes, MK7 6AA, UK \\  \printead{e2}}

\and
\author{\fnms{Paul } \snm{Marriott}\ead[label=e3]{pmarriot@uwaterloo.ca}}
\address{Department of Statistics and Actuarial Science,    University of Waterloo, 200 University Avenue West, Waterloo, Ontario, Canada N2L 3G1 \\\printead{e3}}

\runauthor{Anaya-Izquierdo et al. }

\maketitle

\begin{abstract}
This paper looks at effects, due to the boundary, on inference in logistic regression. It shows that first -- and, indeed, higher -- order asymptotic results are not uniform across the model. Near the boundary, effects such as high skewness, discreteness and collinearity dominate, any of which could render inference based on asymptotic normality suspect. A highly interpretable diagnostic tool is proposed allowing the analyst to check if the boundary is going to have an appreciable effect on standard inferential techniques.
\end{abstract}

\begin{keyword}[class=AMS]
\kwd[Primary ]{62F99}
\kwd[; secondary ]{62-04}
\end{keyword}


  \begin{keyword} 
\kwd{Asymptotic analysis}
\kwd{Boundary effects}
\kwd{Closure}
\kwd{Extended exponential family}
\kwd{Information geometry}
\kwd{Logistic regression}
\kwd{Multinomial distribution}
\end{keyword}

\section{Introduction}

 The fact that the maximum likelihood estimate in a logistic regression model may not exist   is a well-known phenomenon and a number of recent papers have explored its underlying geometrical basis.    \cite{geyer:lik:2009}, \cite{Rina:Fein:Zhou:2009} and  \cite{Rina:Fein:2012} point out that existence, and non-existence, of the  estimate can be fully characterised   by considering the closure of the model as an exponential family. In this formulation it becomes clear that the maximum is always well-defined, but can lie on the boundary rather than in the relative interior. Furthermore, the boundary can be considered as a polytope characterised by  a finite number of extremal points.

This  paper builds on this work and shows that the boundary  affects more than the existence of the maximum likelihood estimate. In particular,  even when the  estimate  exists, the geometry and boundary  can strongly affect inference procedures.   First and  higher order asymptotic results can not be uniformly applied.  Indeed, near the boundary, effects such as high skewness,  discreteness and collinearity dominate, any of which could render inference based on asymptotic normality suspect.   The paper presents a simple diagnostic tool which  allows the analyst to check if the boundary is going to have an appreciable effect on standard inferential techniques. The tool, and the effect that the boundary can have, are illustrated in a well-known example and through   simulated datasets.

\begin{example}\label{Fisher iris} The Fisher iris data set, \cite{Fish:iris:1936}, is often used to illustrate classification and binary regression. Even in this familiar case we show that the boundary  is close enough to have significant effects for inference.  Let us focus on the problem of 
distinguishing the species {\em  Iris setosa} -- coded with $1$ in figures --   from  {\em Iris versicolor} (coded with $0$)  based on  the length of the flower's  sepal.  The left hand panel in Fig.~\ref{simplelogisticfig4} shows the logistic regression fit, while the right hand panel shows a contour plot of the log-likelihood for  intercept parameter $\alpha$ and slope parameter $\beta$. The near singularity of the observed Fisher information is evident.  While this does have a geometric interpretation, this particular collinearity  effect can easily be removed by centring the explanatory variable around its sample mean.  For clarity of exposition we work exclusively with the centred model from now on.

    \begin{figure}[h] 
    \centering
  \includegraphics[width=3in]{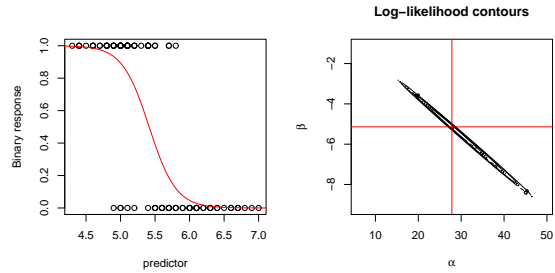}
    \caption{Fisher Iris data example: fit and log-likelhood contours}
    \label{simplelogisticfig4}
 \end{figure}
  
Figure \ref{irislogsiticboundary.eps}, to be discussed   in Section \ref{Examples},  shows various aspects of  sampling distributions under the maximum likelihood fit. These show failure of  first, and higher,  order asymptotics due to   skewness   and discreteness in a number of different ways. These effects can be explained by the closeness of the boundary -- shown in panel (a) as points connected with solid line segments -- in the sequel. In this example, the boundary is just close enough  to start to play a significant role. More extreme examples are shown in  Section \ref{Examples}.

    \begin{figure}[h] 
    \centering
     \includegraphics[width=3in]{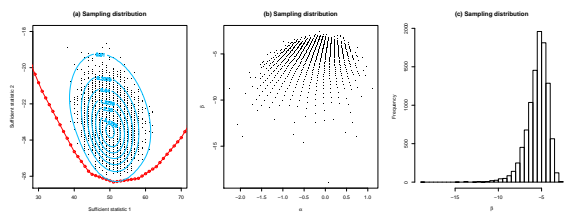}
    \caption{Fisher Iris data example: effects of the boundary on the sampling distribution}
    \label{irislogsiticboundary.eps}
 \end{figure}

\end{example}

\section{Overview of geometry}

This section looks at the  geometry  underlying logistic regression. Section \ref{Extended multinomial models} describes  the more general case of the geometry of extended multinomial models, while Section \ref{Logistic regression} focuses on logistic regression. Section \ref{Limits of first order asymptotics} defines the  diagnostic which allows the analyst to check if the boundary is close enough to substantially effect  first order asymptotic results.  Finally, in Section \ref{Examples},  we return to Example \ref{Fisher iris} and related variants.

\subsection{Extended multinomial models}\label{Extended multinomial models}

The information geometry of the extended multinomial model is considered in \cite{anay:crit:marr:vos:2013a}. The cell probabilities of the extended multinomial  define  the simplex
 \begin{equation}\label{definition of extended multinomial}
\Delta^k:=\left\{{\pi}=(\pi_0, \pi_1,\ldots,\pi_k)^\top\,:\, \pi_i\geq 0\,,\,\sum_{i=0}^k\pi_i=  1\right\}.
\end{equation} The term `extended' refers to the fact that this is the closure of the multinomial model: zero cell probabilities are allowed. The closure of exponential families has been studied by \cite{Barn:1978}, \cite{Brow:1986},   \cite{Laur:1996}  and  \cite{Csis:Matu:2005}.  Of central interest in this paper is the consequence of working in a closed extended family, rather than in the more familiar open parameter space. 

  One important feature -- with obvious implications for first order asymptotics -- is that  the Fisher information can become arbitrarily close to singular as you approach the boundary. This is clearly shown by considering its spectral decomposition. 
With $\pi_{(0)}$  denoting the vector of all  probabilities except $\pi_0$, the Fisher information matrix for the natural (log-odds)  parameters, written as a function of $\pi$,  is the sample size times 
$$I(\pi):= diag(\pi_{(0)}) - \pi_{(0)} \pi_{(0)}^T.$$ Its explicit spectral decomposition is an example of  interlacing eigenvalue results, (see for example  \cite{Horn:John:1985}, Chapter 4).     In particular, suppose $\{\pi_{i}\}_{i=1}^{k}$ comprises $g>1$
distinct values $\lambda_{1}> \cdots >\lambda_{g}>0$, $\lambda_{i}$ occuring
$m_{i}$ times, so that ${\textstyle\sum\nolimits_{i=1}^g} m_{i}=k$. Then, the spectrum of $I(\pi)$ comprises $g$ simple eigenvalues
$\{\widetilde{\lambda}_{i}\}_{i=1}^{g}$ satisfying
\begin{equation}\label{interleaving result}
\lambda_1> \tilde \lambda_1 > \dots > \lambda_g > \tilde \lambda_g \ge 0, 
\end{equation} together, if $g<k$, with $\{\lambda_{i}:m_{i}>1\}$, each such $\lambda_{i}%
$\ having multiplicity $m_{i}-1$.

While the closure of the full multinomial  model  is easy to understand in the representation (\ref{definition of extended multinomial}), the closure of lower dimensional sub-models of $\Delta^k$ expressed in the natural parameters, such as logistic regression models, are more problematic to compute --  although they can be critical  inferentially.  In order to visualise the geometry of the problem of computing limits in exponential families, consider a  low-dimensional example.

\begin{example}   Define   a two dimensional full exponential  subfamily $\pi(\alpha, \beta)$ of $\Delta^3$  where 
$$
\pi(\alpha,  \beta)  \propto \left(\exp\{  \alpha v_{11}+  \beta v_{21}\}, \exp\{  \alpha v_{12}+  \beta v_{22}\}, \exp\{ \alpha v_{13}+  \beta v_{23}\}, \exp\{  \alpha v_{14}+  \beta v_{24}\}  \right)
$$ for  vectors $v_1= (1,2,3,4), v_2=(1,4,9,-1)$. 
Consider directions from the origin $(\alpha, \beta)= (0,0)$ found by writing  $\alpha=\theta \beta$ giving, for each $\theta$,  a   one dimensional full exponential family parameterized by $\beta$ in the direction 
$(\theta +1, 2\theta+4, 3\theta+9, 4\theta- 1  )$.  The aspect of this vector which determines the connection  to the boundary  is the rank  order of its elements, in particular which elements are the maximum and minimum.  For example, suppose the first component was the maximum and  the last the minimum, then as $\beta \rightarrow \pm \infty$ this one dimensional family will be connected to the first and fourth vertex of the embedding four simplex, respectively.  In order to see all possible rank orderings of the components, see the right panel of  Fig.~\ref{boundaryexample.eps} which shows the graphs of the functions  $\{\theta +1, 2\theta+4, 3\theta+9, 4\theta- 1  \}$. The maximum and minimum ranks are determined by the upper and lower envelopes, shown as solid lines.  From this analysis of the envelopes of a set of linear functions, it can be seen that the function $2\theta +4$ is redundant.  It can be shown  that  only three of the four vertices of the ambient 4-simplex have been connected by the model. This is show explicitly in the  left panel of Fig.~\ref{boundaryexample.eps}.

 \begin{figure}[h] 
   \includegraphics[width=3in]{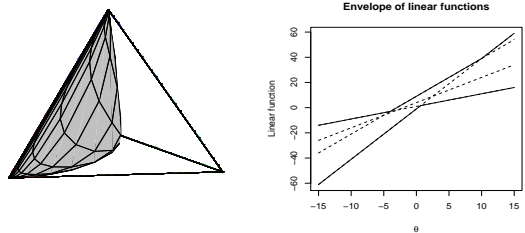}
      \caption{Attaching a two dimensional example to the boundary of the simplex.} 
    \label{boundaryexample.eps}
 \end{figure}
\end{example}

In general,  the problem of finding the limit points in full exponential families inside simplex models is a problem of finding redundant linear constraints. As shown in 
\cite{Edels:Algorthms:1987},  this can be converted, via duality, into the problem of finding extremal points in a finite dimensional affine space.  For an alternative approach, see \cite{geyer:lik:2009}.

\subsection{Logistic regression}\label{Logistic regression}

A logistic regression model  is a full exponential family that   lies in a very high dimensional  simplex when considered as a model for the joint distribution of  $N$ binary response variates. Consider an $N \times D$ design matrix $X$ whose $i^{\rm th}$ row, $x_i^T$, contains the covariate values for the $i^{\rm th}$ case  and a binary response $t \in \{0, 1\}^N$. Let 
$s(a) = \log\left( \frac{a}{ 1- a} \right)$ so that  $s^{-1}(a) = \frac{\exp(a)}{1+\exp(a)}$, the  logistic regression  model being  given by 
$$P(T_i =1)= s^{-1}(x_i^T \beta).$$ 
This is a full exponential family that   lies in the $(2^N-1)$-dimensional simplex when considered as a model for the joint distribution of the $N$ binary response variates. A design matrix  $X$ defines  a $D$-dimensional subset -- an affine subspace in the natural parameters -- and changing the explanatory variates  changes the orientation  of this low dimensional space inside the space of all joint distributions.

\begin{example} Consider a logistic regression with $20$ cases in which $X$ comprises two columns, $1_{20}$, the vector of all ones, and $(1,2, \dots, 20)^T$. It is important to consider the way that this  two-dimensional exponential family is attached to the boundary.  The generalisation of Fig.~\ref{boundaryexample.eps} is shown in  Fig.~\ref{logisticfigenvelope.eps}, where here only lines which are part of the envelope are plotted.   The corresponding vertices which the full exponential family reaches are given by  vectors of the form $z$ with the structure either  $z_i=0$ for $i=1, \dots, h$ and $1$ for $i=h+1, \dots, 20$ or $z_i=1$ for $i=1, \dots, h$ and $0$ for $i=h+1, \dots, 20$. This generalises at once to any single covariate taking distinct values.

 \begin{figure}[h] 
    \centering
       \includegraphics[width=2in]{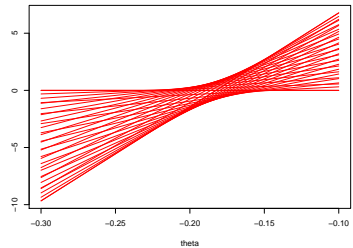}
    \caption{Envelope of lines}
    \label{logisticfigenvelope.eps}
 \end{figure}
 
 \end{example}
 
%
%
 
 \subsection{A  diagnostic tool} \label{Limits of first order asymptotics}
First order asymptotics essentially assumes that the parameter space can be treated as  a Euclidean space with a fixed metric, typically the Fisher information evaluated either at a hypothesised value or the maximum likelihood estimate. This observation allows a simple diagnostic tool  to be developed which gives a sufficient condition for first order methods to be appropriate. If they were appropriate, then the closest point on the boundary should be a large distance, as measured in this fixed metric, from the maximum likelihood estimate -- this length being calibrated using the quantiles of the relevant $\chi^2$ distribution.

When the dimension of the  extended multinomial model is high, relative to sample size, first order asymptotic approximations hold at best on low dimensional subspaces. In the case of logistic regression, consider  the mean parameter space which, with a small but common abuse of notation, we can also consider as the space of  sufficient statistics. In this space, the  vertices which  define the boundary polytope  are also  the extremal points in the convex set of attainable values of the sufficient statistics, see Figs~\ref{irislogsiticboundary.eps} (a) and \ref{simplesimfig5}. In Fig.~\ref{simplesimfig5}, whose details are discussed in Section \ref{Examples},  the contours are defined by the squared distance from the maximum likelihood estimate relative to the Fisher information there. The largest contour is calibrated by the $99\%$-quantile of the  $\chi^2_2$ distribution. The diagnostic is simply based on checking if this contour crosses the boundary or not. If it does cross the boundary, for example  marginally in the left panel and strongly in the right, then we regard first order asymptotic normality as suspect.

We note in passing that, in a general multinomial model setting, distances to the boundary can be easily computed, via quadratic programming, and in fact have a closed form. Let $Q_{\pi_0}(\pi)$ be the squared distance from $\pi_0$ to $\pi$ measured with respect to the Fisher information at $\pi_0$. Using this distance function, the squared distance to the face defined by the index set $I$ from the point  $\pi_0$  is  
$$
Q_{\pi_0}(\pi)=\frac{ \pi_I}{1-  \pi_I},
$$ where $ \pi_I= \sum_{i \in I}  \pi_{0\,i}$.

\section{Examples and discussion}\label{Examples} 

Let us return to Example \ref{Fisher iris}.   Consider first panel (a) of Fig.~\ref{irislogsiticboundary.eps}. This is a plot of randomly sampled  sufficient statistics  -- plotted by dots --  generated from the distribution identified by the maximum likelihood estimate. The vertices in the extended multinomial model to which the logistic model is connected correspond to  a set of points in the sample space -- plotted with circles -- and the edges which connect these points  define a one dimensional boundary --  plotted with straight lines. The image of the boundary is a polytope and it is clear that for the iris data this sample is getting very close to the boundary.  

In the example the boundary is just close enough for the first order approximations  to start to break down. This is happening in a number of ways. There is noticeable discreteness in the sample, each vertical streak corresponding to one of the extremal points on the boundary. This effect becomes more pronounced the closer the boundary becomes, as shown below. 

The second way that the first order approximation breaks down is that the effect of higher order terms in the asymptotic expansions are starting to make themselves felt.  This is illustrated by the solid contour lines, which are defined by the two dimensional  Edgeworth expansion of the sampling distribution, see \cite{Barn:Cox:1989}.  It can be seen that these are not centred ellipses, as would have been expected if the normal approximation was adequate. Rather a distortion caused by the boundary is becoming evident. 

Panel (b) in Fig.~\ref{irislogsiticboundary.eps} shows the sampling distribution of the maximum likelihood estimate. Near the boundary the very high degree of non-linearity between the maximum likelihood estimates and the natural sufficient statistics  has a very strong effect, as can be seen. The strong directional features correspond to directions of recession as described in  \cite{geyer:lik:2009}.   The effect of this non-linearity can be further seen in panel (c) which plots the marginal distribution of $\widehat \beta$, the estimate of the slope parameter, whose large skewness clearly indicates non-normality.

\begin{figure}[h] 
    \centering
      \includegraphics[width=3in]{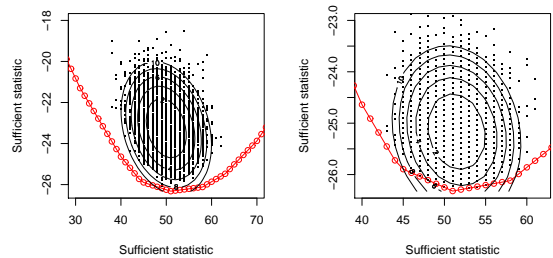}
    \caption{Distribution of sufficient statistics near the boundary}
    \label{simplesimfig5}
 \end{figure}

In order to show how all these aspects become stronger when the boundary gets closer, consider Fig.~\ref{simplesimfig5}. The left hand panel shows the same information as in  Fig.~\ref{irislogsiticboundary.eps}, but now with the contours of our proposed diagnostic plotted. This is the case where the maximum likelihood estimate from Example \ref{Fisher iris} is used and, as can be seen the diagnostic line, calibrated by the $99\%$ quantile of the $\chi^2_2$ distribution, just touches the boundary. This implies that boundary effects are starting   to affect inference, as described above.
 
 In the right hand panel of Fig.~\ref{simplesimfig5}, the sampling has been done from a distribution much closer to the boundary. As can be clearly seen, the diagnostic curves strongly cut the boundary. The effects on inference in this case are much stronger. The discretisation effects are much stronger;  in particular, note that  most of the probability mass on the boundary lies on a relatively small number of vertices -- plotted as circles in the figure.   The effect on the sampling distribution of the maximum likelihood estimate is even more extreme. There is now an appreciably large probability that a sampled vector of sufficient statistics  lies on a boundary point, which implies an `infinite' slope estimate. This gives very strong departures from normality for both the joint and marginal distributions.

%
\bibliographystyle{plain} 
\bibliography{logisticgeometry} 

\begin{thebibliography}{10}

\bibitem{anay:crit:marr:vos:2013a}
K.~Anaya-Izquierdo, F.~Critchley, P.~Marriott, and P.~W. Vos.
\newblock Computational information geometry: foundations.
\newblock {\em Submitted to Geometric Science of Information 2013}, 2013.

\bibitem{Barn:1978}
O.~E. Barndorff-Nielsen.
\newblock {\em Information and exponential families in statistical theory}.
\newblock John Wiley \& Sons, 1978.

\bibitem{Barn:Cox:1989}
O.~E. Barndorff-Nielsen and D.~R. Cox.
\newblock {\em Asymptotic techniques for use in statistics}.
\newblock Chapman \& Hall, 1989.

\bibitem{Brow:1986}
L.~D. Brown.
\newblock {\em Fundamentals of statistical exponential families: with
  applications in statistical decision theory}.
\newblock Institute of Mathematical Statistics, 1986.

\bibitem{Csis:Matu:2005}
I.~Csiszar and F.~Matus.
\newblock Closures of exponential families.
\newblock {\em The Annals of Probability}, 33(2):582--600, 2005.

\bibitem{Edels:Algorthms:1987}
H.~Edelsbrunner.
\newblock {\em Algorithms in combinatorial geometry}.
\newblock Springer-Verlag: NewYork, 1987.

\bibitem{Rina:Fein:2012}
S~Fienberg and A.~Rinaldo.
\newblock Maximum likelihood estimation in log-linear models: Theory and
  algorithms.
\newblock {\em Annals of Statist.}, 40:996--1023, 2012.

\bibitem{Fish:iris:1936}
R.~A. Fisher.
\newblock The use of multiple measurements in taxonomic problems.
\newblock {\em Annals of Eugenics}, 7(2):179--188, 1936.

\bibitem{geyer:lik:2009}
C.~J. Geyer.
\newblock Likelihood inference in exponential families and directions of
  recession.
\newblock {\em Electron. J. Statist.}, 3:259--289, 2009.

\bibitem{Horn:John:1985}
R.~A. Horn and C~.R. Johnson.
\newblock {\em Matrix Analysis}.
\newblock CUP, 1985.

\bibitem{Laur:1996}
S.~L. Lauritzen.
\newblock {\em Graphical models}.
\newblock Oxford University Press, 1996.

\bibitem{Rina:Fein:Zhou:2009}
A.~Rinaldo, S.~Fienberg, and Y.~Zhou.
\newblock On the geometry of discrete exponential families with applications to
  exponential random graph models.
\newblock {\em Electron. J. Statist.}, 3:446--484, 2009.

\end{thebibliography}

\end{document}